\documentclass[aps,twocolumn,amsmath,amssymb,prl,showpacs]{revtex4}
\usepackage[colorlinks=true,citecolor=red,urlcolor=blue]{hyperref}
\usepackage{amsfonts}
\usepackage{graphicx}
\begin{document}
\newcommand{\upd}{\mbox{d}}
\newcommand{\chem}[1]{{ #1}}

\title{Current driven magnetization dynamics in ferromagnetic nanowires
with Dzyaloshinskii-Moriya interaction}

\author{O. A. Tretiakov }
\email{olegt@tamu.edu}
\author{Ar.~Abanov}

\affiliation{
            Department of Physics,
            MS 4242,
	    Texas A\&M University,
            College Station, TX 77843-4242, USA
}

\date{December 22, 2009}

\begin{abstract}
We study current induced magnetization dynamics in a long thin
ferromagnetic wire with Dzyaloshinskii-Moriya interaction (DMI).  We
find a spiral domain wall configuration of the magnetization and
obtain an analytical expression for the width of the domain wall as a
function of the interaction strengths.  Our findings show that above a
certain value of DMI a domain wall configuration cannot exist in the
wire. Below this value we determine the domain wall dynamics for small
currents, and calculate the drift velocity of the domain wall along
the wire.  We show that the DMI suppresses the minimum value of
current required to move the domain wall.  Depending on its sign, the
DMI increases or decreases the domain wall drift velocity.
\end{abstract}


\maketitle

\textit{Introduction.}  A number of recent experiments, performed in
various metallic magnets, have shown the spiral structure of
magnetization due to Dzyaloshinskii-Moriya interaction (DMI)
\cite{Pfleiderer01, Doiron-Leyraud03, Pfleiderer04, Uchida06,
  Uchida08, Meckler09, Ferriani08, Bode07}.  In particular, the B20
structure of ferromagnets, such as \chem{MnSi}, which lacks strict
space-inversion symmetry, leads to a long-wavelength helical twist in
the magnetization \cite{Pfleiderer01,Doiron-Leyraud03,Pfleiderer04}.
Furthermore, the direct space-time observation of the spiral structure
by Lorentz microscopy became possible for DMI-induced helimagnets
\cite{Uchida06,Uchida08}.  Using spin-polarized scanning tunneling
microscopy it has been shown that the magnetic order of 1 monolayer
\chem{Mn} on \chem{W}(001) is a left-handed spiral \cite{Ferriani08}
and that the magnetic structure of the \chem{Fe} double layer on
\chem{W}(110) is a right-rotating spiral \cite{Meckler09}.  All these
spiral states are the consequence of DMI.

A spin-polarized current flowing through such spiral magnetic
structures would exert a spin-torque which could be used for
manipulations of the magnetization with potential applications. For
example, in magnetic memory devices \cite{Parkin08,Hayashi08} the key
issue is to manipulate the domain wall (DW) configurations by means of
magnetic fields and/or spin-polarized current. Therefore,
current-induced dynamics of spiral magnets is an important subject of
technological relevance.

One of the most important factors which effects the DW motion is
pinning. The DW pinning can have ``extrinsic'' and ``intrinsic''
nature. The extrinsic pinning is due to surface roughness and other
irregularities of the wires which brake translational invariance. On
the other hand, the intrinsic pinning is present even in ideally
smooth (translation invariant) nanowires. It depends on the wire
geometry and material parameters which can be described by
anisotropies. Although extrinsic pinning can be significantly reduced
in the near future with the help of more sophisticated wire
fabrication techniques, the intrinsic pinning is always present.
Therefore, in this Letter we concentrate on the more important case of
DW dynamics with the intrinsic pinning.

We determine the effect of a polarized current on the magnetization
configuration in the ferromagnetic wire with both strong easy-axis
anisotropy along its axis and weak anisotropy in the plane transverse
to the wire.  The DMI, which arises from spin-orbit scattering of
electrons in non-centrosymmetric magnetic materials is typically
irrelevant in bulk metals as their crystals are
inversion-symmetric. However, in low-dimensional systems (such as
atomic layers and wires), which lack structural inversion symmetry,
the DMI in the presence of softened ferromagnetic exchange coupling
leads to the formation of the spiral spin structures.

The main goal of this Letter is to study the influence of DMI on the
magnetization dynamics in ferromagnets.  We obtain the expression for
the DW width as a function of the DMI constant, uniaxial anisotropy
along the wire, and exchange interaction constant. We find that there
is a critical value of the DMI above which a DW configuration cannot
exist in the wire. This result can have an important implication for
the future experiments by setting a limit on the devices with DMI
which use DWs for information manipulation.  Below this critical value
of DMI the DW can propagate along the wire and rotate around its
axis. Any angle is equally favorable for the DW if there is no
anisotropy in the transverse plane. Generally speaking in most wires
there exists such an anisotropy due to the asymmetry of the wire cross
section. We show that it leads to a chosen direction of the
magnetization in the center of the DW, so that the wall cannot rotate
freely anymore. Therefore, if a polarized current is passing through
such a wire, the DW will move only if the current is larger than a
certain critical value. This value corresponds to the minimal torque
needed to be pumped into the system to rotate the spins of the DW
around the wire's axis.

We investigate the dynamics of the DW in the small transverse
anisotropy regime.  In particular, we find the drift (average)
velocity of the DW in the wire with DMI. Our findings also show that
DMI decreases the critical value of current required to move a DW.  To
obtain all these results for the DW dynamics, we use a universal
method for finding zero mode dynamics of spin textures.  This method
is described in detail in the supplementary material~\cite{supp1}.

\begin{figure}
\includegraphics[width=1\columnwidth]{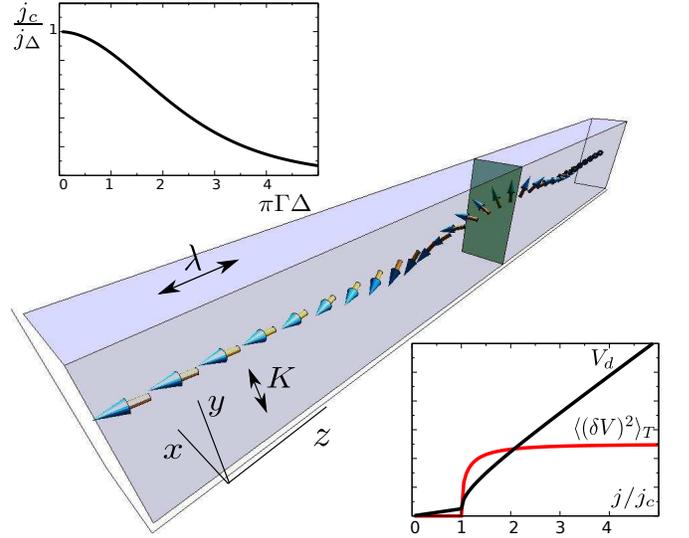} 
\caption{(color online) Sketch of the wire with magnetization profile
  for a DW.  $\lambda$ and $K$ denote the anisotropies along and
  transverse to the wire, respectively. The upper inset shows the
  dependence of $j_{c}$ on the twist $\Gamma \Delta $,
  Eq.~\eqref{eq:jcbeta}; the lower inset shows drift velocity $V_{d}$
  and variance $\langle (\delta V)^{2}\rangle_{T}$ (in arb. units)
  vs. current $j$, see Eqs.~\eqref{eq:Vd} and \eqref{eq:dVd}.}
\label{fig:wire} 
\end{figure}

\textit{Model.}  We employ a simple theoretical model of a ferromagnet
with DMI and anisotropies which highlights a new kind of behavior of
DW structures.  We consider a Hamiltonian for a ferromagnet which has
two terms describing the exchange and Dzyaloshinskii-Moriya
interactions \cite{Dzyaloshinskii58,Moriya60}.  Without anisotropies
in the continuous limit this Hamiltonian takes the form,
\begin{equation} 
\mathcal{H}_0=\int
  d^{3}r\left[\frac{J_{0}}{2}\left(\mathbf{\nabla}\mathbf{M}\right)^{2}
+D_{0}\mathbf{M}\cdot\left(\mathbf{\nabla}\times\mathbf{M}\right)\right].
\label{eq:HamG}
\end{equation}
Here $\mathbf{M}$ is a magnetization vector, $J_{0}>0$ is exchange
interaction constant, and $D_{0}$ is the DMI constant. We study a
ferromagnetic wire which is modeled as a one-dimensional (1D)
classical spin chain~\footnote{The width of the wire is much smaller
  then any characteristic length of the magnetic structure, but is
  large enough so that the total spin in a cross section is large.},
where the wire is along the $z$-axis, see Fig.~\ref{fig:wire}. For the
thin long wire with uniaxial anisotropy Hamiltonian \eqref{eq:HamG}
modifies to
\begin{equation}
\mathcal{H}=\int
\upd z\left[\frac{J}{2}\left(\partial\mathbf{S}\right)^{2}
+D\mathbf{S}\cdot\left(\mathbf{e}_{z}\times
\partial\mathbf{S}\right)-\lambda S_z^{2} \right].
\label{eq:Ham}
\end{equation}
Here $\mathbf{e}_{z}$ is the unit vector in $z$ direction,
$\partial=\partial/\partial z$, and we introduced normalized
magnetization vector $\mathbf{S}=\mathbf{M}/M$, so that
$\mathbf{S}^{2}=1$, $D=D_{0}/(AM^{2})$, and $J=J_{0}/(AM^{2})$, where
$A$ is the cross-sectional area of the wire.  The last term in
Eq.~\eqref{eq:Ham} is due to uniaxial anisotropy (with the anisotropy
constant $\lambda= \lambda_0/(AM^{2})$) which shows that the system
favors the magnetization along the wire.

To study the magnetization dynamics we employ the generalized
Landau-Lifshitz-Gilbert equation \cite{Zhang04,Thiaville05} for 1D
wire with current $j$ along the wire:
\begin{equation}
  \dot{\mathbf{S}}=\mathbf{S}\times\mathbf{H}_{e}
-j\partial\mathbf{S}
+\beta j\mathbf{S}\times\partial\mathbf{S}
+\alpha\mathbf{S}\times\dot{\mathbf{S}}.\label{eq:LLG}
\end{equation}
where $\mathbf{H}_{e}=\delta\mathcal{H}/\delta\mathbf{S}$,
$\dot{\mathbf{S}}=d\mathbf{S}/dt$, $\alpha=\alpha_{0}/M^{2}$ and
$\alpha_{0}$ is the Gilbert damping constant, $\beta=\beta_{0}/M^{2}$
and $\beta_{0}$ is the constant of nonadiabatic current term, time is
measured in the units of the gyromagnetic ratio
$\gamma_{0}=g\left|e\right|/(2mc)$, and the current $j$ is measured in
units of $a^{3}/(2eM\gamma_{0})$ where $a$ is the lattice constant.
Generally speaking one also has to specify the boundary conditions for
Eq. \eqref{eq:LLG}.

A general solution of one-dimensional LLG equation \eqref{eq:LLG} can
always be presented in the form
\begin{equation} 
\partial\mathbf{S} =
\Gamma(z,t)\mathbf{e}_{z}\times\mathbf{S}
+\Lambda(z,t)\mathbf{S}\times[\mathbf{e}_{z}\times\mathbf{S}],
\label{eq:GSol}
\end{equation}
where $\Gamma$ and $\Lambda$ are in general two independent functions
of $z$ and $t$; it also follows that $\partial S_{z}=
\Lambda(1-S_{z}^{2})$.

\textit{Zero current.}  First we consider the simplest case of zero
current ($j=0$) and look for a time-independent magnetization
configuration. This means that we need to minimize Hamiltonian
\eqref{eq:Ham} which can be written up to a constant in the form
\begin{equation} 
\mathcal{H}\!=\!\!\int\!\!
\upd z\left[\frac{J}{2}\left(\partial\mathbf{S}-\frac{D}{J}\mathbf{e}_{z}
\times\mathbf{S}\right)^{2}\!\!\!
+\!\left(\lambda-\frac{D^{2}}{2J}\right)(1-S_{z}^{2})\right]\!.
\label{eq:HamNewForm}
\end{equation}
The spin configuration depends on the sign of
$\lambda-D^{2}/2J$.

For $2J\lambda<D^{2}$ the minimum of the second term is at
$S_{z}=0$. The first term is minimized by the condition
$\partial\mathbf{S}=\frac{D}{J}\mathbf{e}_{z}\times\mathbf{S}$, so
that the solution is a spiral,
\begin{equation}
\mathbf{S}=(\cos(\Gamma z+\phi_{0}),\sin(\Gamma z+\phi_{0}),0)^{T},
\quad \Gamma =D/J.
\label{eq:spiral}
\end{equation}
The ground state is thus unique and there is no DW configuration.
Therefore, for the wires with weak enough uniaxial anisotropy and/or
exchange constant the spiral magnetization state can prevent the
formation of DWs.

For $2J\lambda>D^{2}$ the minimum of the second term is at
$S_{z}=\pm1$. This also minimizes the first term in
Eq.~\eqref{eq:HamNewForm}.  Thus, $S_{z}=\pm1$ are the two solutions,
and a DW can exist in the wire as a transition from one solution to
another.

Below we study the statics and dynamics of such a DW in the wire, and
therefore we concentrate on the case $2J\lambda>D^{2}$.  Then the
boundary conditions for Eq. \eqref{eq:LLG} are $S_{z}\rightarrow \pm
1$ at $z\rightarrow \pm \infty $.  To find the static configuration of
the DW we consider the solution in the form~\eqref{eq:GSol}.
Substituting it into Hamiltonian~\eqref{eq:HamNewForm}, we find
\begin{equation}
\mathcal{H}=\!\int\!\! \upd z\left[\frac{J}{2}\left(\Gamma
-\frac{D}{J}\right)^{2}+\frac{J}{2}\Lambda^{2}
+\lambda-\frac{D^{2}}{2J}\right](1-S_{z}^{2}).
\label{eq:HamGammaDelta}
\end{equation}
The minimization of the first term sets 
\begin{equation}
\Gamma=D/J\,,\label{eq:Gamma}
\end{equation}
cf. Eq.~\eqref{eq:spiral}.  Using parametrization $S_{z}=\tanh f(z)$,
we obtain
\begin{equation}
\mathcal{H} = \frac{J}{2}\!\int\!\! \upd z\frac{(\partial f)^{2}
+\Delta^{-2}}{\cosh^{2}f}\,,
\quad
\Delta^{-2} =\Delta_{0}^{-2}-\Gamma^{2},
\label{eq:Hamf}
\end{equation}
where $\Delta_{0}^{2}=\sqrt{J/2\lambda }$ is the DW width in the
absence of DMI. The straightforward minimization of
Eq.~\eqref{eq:Hamf} gives $f=z/\Delta$ or $\Lambda=1/\Delta$, and in
components the solution takes the form
\begin{subequations}
\begin{eqnarray}
&&S_{x}=\frac{\cos(\Gamma
    (z-z_{0})+\phi)}{\cosh((z-z_{0})/\Delta)},\\
&&S_{y}=\frac{\sin(\Gamma
    (z-z_{0})+\phi)}{\cosh((z-z_{0})/\Delta)},\\
&&S_{z}=\tanh(z-z_{0})/\Delta),
\end{eqnarray}
\label{eq:domainSolutionFull}
\end{subequations}
where the angle $\phi$ is the tilt of the DW, and $z_{0}$ is its
position (both arbitrary). We see that $2\pi/\Gamma $ is the pitch of
the spiral, $\Delta$ is the width of the DW and $\Gamma \Delta $ is
thus the twist of the DW.  Both $\Gamma$ and $\Delta$ have the same
functional dependencies in terms of $J_0$, $D_0$, and $\lambda_0$ as
in terms of $J$, $D$, and $\lambda$.  According to its definition in
Eq.~\eqref{eq:Hamf}, $\Delta$ becomes infinite in the limit
$2J\lambda=D^{2}$ and DW cannot be sustained in the wire.

The energy of the DW is $E=2J/\Delta=2\sqrt{2J\lambda -D^{2}}$. It
vanishes when $D^{2}$ approaches $2J\lambda $.

The $z$ component of the magnetization \eqref{eq:domainSolutionFull}
is the same as that of a standard (without DMI) DW of width $\Delta $.
The direction of the twist of the DW depends on the sign of DMI and
can be either clock or counterclockwise.

Parameters $z_{0}$ and $\phi$ in Eq.~\eqref{eq:domainSolutionFull}
correspond to two zero-modes of the system. These modes are the most
relevant if the system is perturbed. The time-dependent solution then
can be represented in the form of a moving and rotating DW plus a
small correction to its shape. The requirement that the correction
remains small during the motion leads to the equations for the
velocity and angular velocity of the DW.  A detailed derivation of
these equations is presented in the supplementary material
\cite{supp1}. Below we present the results and discuss their
implications.

\textit{Small currents.}  First we find the magnetization dynamics in
the wire for small applied currents. We denote the solution
\eqref{eq:domainSolutionFull} for the DW without a current, as
$\mathbf{S}_{0}(z)$.  When the current is applied we expect the DW to
move and rotate.  The full dynamics is described by the equation
\begin{equation}
\label{eq:LLgh}
\dot{\mathbf{S}}=\mathbf{S}\times\mathbf{H}_e +\mathbf{h},
\qquad \mathbf{H}_e=\delta\mathcal{H}/\delta\mathbf{S},
\end{equation}
where the correction $\mathbf{h}$ for small currents is
$\mathbf{h}=\mathbf{h}_{j}$,
\begin{equation}
\mathbf{h}_{j}=-j\partial\mathbf{S}_{0}+(\beta
j-\alpha\dot{z}_{0})\mathbf{S}_{0}\times\partial\mathbf{S}_{0}
+\alpha\dot{\phi}\mathbf{S}_{0}\times\mathbf{e}_{z}\times\mathbf{S}_{0}.
\label{eq:Halpha}
\end{equation}
This correction gives the following results for the DW velocity and
angular velocity:
\begin{equation}
\dot{z}_{0}
=\frac{1+\alpha\beta+(\alpha-\beta)\Gamma\Delta}{1+\alpha^{2}}j,\quad 
 \dot{\phi}=\frac{(\alpha-\beta)\Delta}{(1
+\alpha^{2})\Delta_{0}^{2}}j\,.
\label{eq:phidot}
\end{equation}
A few conclusions can be made from these equations. 

i.) The direction of the DW rotation depends only on the relative
strength of the two dissipative terms in the LLG
Eq.~\eqref{eq:LLG}. Remarkably, the sign of the DMI correction to the
DW velocity depends on weather the DW rotates in the same direction as
the twist of the DW.

ii.) For $\beta=0$ there is an integral of motion $\alpha
(\Gamma^{2}+\Delta^{-2})z_{0} -(1/\Delta +\alpha\Gamma)\phi
=\text{const}$.  If we take $1/\Delta=0$ which corresponds to a
perfect spiral state (DW width is infinite), this invariant just
describes the rotation of the spiral while it moves.

iii.) At very large twists $\Gamma \Delta $, $\dot{z}_{0} =\dot{\phi}
\Gamma J/2\lambda $, independently of both $\alpha$ and $\beta $.

iv.) The DW rotation and its velocity diverge when $D^{2}$ approaches
$2\lambda J$. This nonphysical result is the consequence of the fact
that our derivation of Eq.~ \eqref{eq:phidot} neglects all modes
except the zero ones. However, in the limit of $D^{2}\rightarrow
2\lambda J$ the breathing mode (the mode that corresponds to the
change of the DW width and pitch) softens and its dynamics cannot be
neglected~\cite{future}.

In line with the general result \cite{Barnes05}, in the special case of
$\alpha =\beta $, the DW does not rotate and just moves with the
velocity which depends on current only.

\textit{Small anisotropy in the transverse plane.}  In order to
account for the anisotropy in the transverse plane we introduce a
correction to Hamiltonian \eqref{eq:Ham} in the form
\begin{equation} 
\mathcal{H}_{xy}=\int \upd zKS_{y}^{2}(z),
\label{eq:xy}
\end{equation} 
where the anisotropy constant $K>0$ is very small. 

The presence of this anisotropy fixes the tilt angle $\phi$ of the
solution \eqref{eq:domainSolutionFull}. To show it we calculate the
correction to the energy to the first order in $K$ by substituting Eq.
\eqref{eq:domainSolutionFull} into Eq.~\eqref{eq:xy}.  Assuming the
wall to be at the origin ($z_{0}=0$), we obtain $\delta_{1}E = K\Delta
-\frac{2\pi K\Gamma\Delta^{2}}{\sinh(\pi\Gamma\Delta)}\cos(2\phi) $.
This correction has a minimum at $\phi=0, \pi$. When DMI is absent
($\Gamma=0$) this correction reduces to $K\Delta [1-2\cos(2\phi)]$.

\textit{Dynamics and transverse anisotropy.}  Now we find how the
small anisotropy in the transverse plane affects the magnetization
dynamics. The correction $\mathbf{h}$ defined in Eq.~\eqref{eq:LLgh}
takes the form $\mathbf{h}=\mathbf{h}_{j}+\mathbf{h}_{xy}$, where
$\mathbf{h}_{j}$ is given by Eq.~\eqref{eq:Halpha} and
\begin{equation}
\mathbf{h}_{xy} =\mathbf{S}\times\frac{\delta\mathcal{H}_{xy}}{\delta\mathbf{S}}
= 2KS_{y}\mathbf{S}\times\mathbf{e}_{y}\, .
\label{eq:vecHxy}
\end{equation}

This perturbation leads to the following equations for the position of
the DW and the tilt angle:
\begin{eqnarray}
\dot{z}_{0} &=& \frac{\beta}{\alpha}j
+\frac{(\alpha-\beta)(1+\alpha\Gamma\Delta)}{\alpha(1+\alpha^{2})}
\left[j-j_{c}\sin(2\phi)\right],
\label{eq:AdotZ0}\\
\dot{\phi} &=&\frac{(\alpha-\beta)
\Delta}{(1+\alpha^{2})\Delta_{0}^{2}}\left[j-j_{c}\sin(2\phi)\right],
\label{eq:AdotPhi}
\end{eqnarray}
where the critical current $j_{c}$ is given by
\begin{equation}
j_{c}=j_{\Delta }\frac{\pi \Gamma\Delta}{\sinh(\pi\Gamma\Delta)},
\qquad j_{\Delta}=\frac{\alpha K\Delta }{\left|\alpha-\beta\right|}\,.
\label{eq:jcbeta}
\end{equation}
$j_{\Delta }$ is a critical current for the domain wall of the same
width, but without the twist.  These equations reduce to
Eq.~\eqref{eq:phidot} for $K=0$.  We also note that
Eq.~\eqref{eq:jcbeta} is correct only in the first order in $K$.

The critical current $j_{c}$ is \textit{exponentially suppressed} for
twists $\Gamma \Delta \gtrsim 1/\pi $.  For small twists $j_{c}\approx
j_{\Delta }(1-\pi^{2}(\Gamma \Delta )^{2}/6)$.  Note that $j_{c}$ in
Eq.~\eqref{eq:jcbeta} diverges at $\alpha=\beta$, that is the DW does
not spin for any current \cite{Barnes05}.

For $j<j_{c}$ the DW tilts by the angle $\sin (2\phi_{j})=j/j_{c}$ and
moves with a constant velocity $\dot{z}_{0}=j\beta /\alpha $, if
$\beta =0$, the DW does not move at all.  For $j>j_{c}$ the DW both
spins and moves along the wire.

\begin{figure}
\includegraphics[width=1\columnwidth]{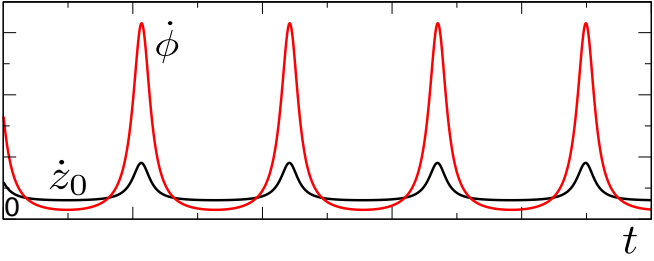} 
\caption{(color online) Velocities $\dot{z}_{0}$ and $\dot{\phi }$ given by
  Eqs.~\eqref{eq:AdotZ0} and \eqref{eq:AdotPhi} at $j=1.1j_{c}$
  vs. time. (In arb. units) }
\label{fig:v} 
\end{figure}

Eqs.~\eqref{eq:AdotPhi} and \eqref{eq:AdotZ0} can be solved
analytically. The solution gives both the velocity and angular
velocity, which periodically depend on time~\cite{supp2} (see
Fig.~\ref{fig:v}), with the period $T$ and average angular velocity
$\Omega$ given by \cite{supp1}:
\begin{equation}
\Omega = \frac{2\pi }{T}=\frac{(\alpha-\beta)\Delta}{(1+\alpha^{2})\Delta_{0}^{2}}\sqrt{j^{2}-j_{c}^{2}}.
\label{eq:period}
\end{equation}
More experimentally relevant, however, is the average (drift) velocity
of the DW $V_{d}=\langle \dot{z}_{0}\rangle_{T}$. For any current it
is given by \cite{supp1}
\begin{equation}
V_{d}=\left\{ 
\begin{array}{ll}
\!\!\!\frac{\beta}{\alpha}j, &\!\!\!\!\quad\mbox{for \ensuremath{j<j_{c}},}\\
\!\!\!\frac{\beta}{\alpha}j+\frac{(\alpha-\beta)(1
+\alpha\Gamma\Delta)}{\alpha(1+\alpha^{2})}
\sqrt{j^{2}-j_{c}^{2}}, &\!\!\!\!\quad\mbox{for \ensuremath{j>j_{c}}}.
\end{array}\right.
\label{eq:Vd}
\end{equation}
The square of the deviation of the velocity from the drift velocity,
Eq.~\eqref{eq:Vd}, $\langle (\delta V)^{2}\rangle_{T}$ is
\begin{equation}
\label{eq:dVd}
\langle (\delta V)^{2}\rangle_{T}=\left\{ 
\begin{array}{ll}
0, &\mbox{for \ensuremath{j<j_{c}},}\\
\left[\frac{(\alpha-\beta)(1+\alpha\Gamma\Delta)}{\alpha(1+\alpha^{2})}j_{c} 
\right]^{2}\frac{\sqrt{j^{2}-j_{c}^{2}}}{j+\sqrt{j^{2}-j_{c}^{2}}}, 
&\mbox{for \ensuremath{j>j_{c}}}.
\end{array}\right.
\end{equation} 
Both $V_{d}(j)$ and $\langle (\delta V)^{2}\rangle_{T}$ are shown in
the inset of Fig.~\ref{fig:wire}.  For large currents, $j\gg j_{c}$,
the drift velocity asymptotically approaches the velocity given by
Eq.~\eqref{eq:phidot}, while $\langle (\delta V)^{2}\rangle_{T}$
approaches a constant.

\textit{Summary.}  We have studied the effects of DMI on the
magnetization statics and dynamics in a thin ferromagnetic wire. We
have derived a simple criterion which determines whether the wire with
the spiral magnetization state can sustain a DW configuration. Namely,
in the wires with weak enough uniaxial anisotropy and/or exchange
constant compared to DMI constant ($2J\lambda<D^{2}$) a DW cannot be
formed.  In the opposite case ($2J\lambda>D^{2}$) we have found the
spiral magnetization state with a DW in the wire.  For $\beta=0$ the
wall moves along the wire only if the applied current is above $j_{c}$
given by Eq.~\eqref{eq:jcbeta}. The variance of the velocity in this
regime is given by $\langle (\delta V)^{2}\rangle_{T}
=V_{d}^{2}(j/\sqrt{j^{2}-j_{c}^{2}}-1 )$.  For $\beta\neq0$ the DW
moves but does not rotate for currents below $j_{c}$.  Above
$j_{c}$ the DW both moves and rotates \cite{supp2}. Our result,
Eq.~\eqref{eq:jcbeta}, shows that the critical value of current is
suppressed by DMI. We also have derived the expression,
Eq.~\eqref{eq:Vd}, for the drift velocity $V_{d}$ of the DW for all
values of current.  It shows that above the critical current $j_{c}$
the drift velocity can be enhanced by DMI.

We believe that our findings can be experimentally observed, e.g.,
with the use of the scanning tunneling microscopy which was employed
to reveal the DW structure in ultrathin \chem{Fe} nanowires \cite{
  Kubetzka03, Vedmedenko04}.  We note that in a realistic experimental
setting besides the ``intrinsic'' pinning there always going to be an
extrinsic pinning due to a nonideal shape of the wire. It is, however,
clear that in the near future the development of better
nanofabrication techniques will lead to the situation when one has to
worry mostly about the ``intrinsic'' effect.

We are grateful to J.~Sinova and Yu.~Adamov for valuable discussions.
This work was supported by the NSF Grant No. 0757992 and Welch
Foundation (A-1678).

\bibliography{magnetizationDynamics}
\end{document}